\documentclass[doublespacing]{elsart}

\usepackage{epsfig}

\usepackage{amssymb}

\begin{document}

\begin{frontmatter}

\title{Statistical  properties of  pinning fields  in  the 3d-Gaussian
RFIM}

\author{Xavier   Illa  \corauthref{cor}},  
\corauth[cor]{Corresponding   author:  Xavier  Illa,   Dept.   E.C.M.,
Universitat de Barcelona, Facultat de F\'{\i}sica, Diagonal 647, 08028
Barcelona,   Catalonia  (Spain)  \\   e-mail:  xit@ecm.ub.es   ;  FAX:
34934021174  \\  {\bf  Session  code:  FC-06, Paper  code:
027X30I01Y} }
\author{Eduard Vives}

\address{Departament  d'Estructura  i  Constituents de  la  Mat\`eria,
Universitat  de Barcelona,  Diagonal 647,  08028  Barcelona, Catalonia
(Spain).}

\begin{abstract}
We have defined  pinning fields as those random  fields that keep some
of the magnetic moments unreversed  in the region of negative external
applied field  during the demagnetizing  process.  An analysis  of the
statistical properties of such  pinning fields is presented within the
context of the Gaussian Random  Field Ising Model (RFIM). We show that
the average of the pinning fields exhibits a drastic increase close to
the coercive field and that  such an increase is discontinuous for low
degrees of  disorder.  This behaviour  can be described  with standard
finite size scaling (FSS) assumptions.  Furthermore, we also show that
the pinning fields corresponding to states close to coercivity exhibit
strong statistical correlations.
\end{abstract}

\begin{keyword}
hysteresis \sep random field ising model \sep pinning fields  

\PACS 75.60.Jk \sep 75.10.Nr \sep 75.10.Hk
\end{keyword}
\end{frontmatter}

\newpage

\section{Introduction}
\label{Introduction}
The  magnetization reversal  process  in a  ferromagnet  is a  complex
dynamical   process   which    is   still   not   totally   understood
\cite{Bertotti1998} .   It is,  nevertheless, very important  for both
fundamental and technological reasons.   Different factors play a role
in determining the metastable path  that the system follows to reverse
the  magnetization  from full  positive  saturation  to full  negative
saturation  when  the  external  field is  decreased.   Among  others,
thermal fluctuations, long range  dipolar forces, anisotropy and local
forces  due to  disorder,  compete  together in  order  to decide  the
sequence of magnetic domains that transform.

A first simplification  of the problem consist in  neglecting the role
of  fluctuations  and relaxation  effects.   This  corresponds to  the
limiting  case   of  ``rate-independent''  hysteresis.   Magnetization
reversal  steps  occur  as  almost  instantaneous  avalanches  joining
metastable states.   Such avalanches  are triggered when  the external
forces induced by the applied  field are strong enough to overcome the
internal  energy  barriers  which   are  caused  by  exchange  forces,
long-range dipolar forces and forces created by disorder.

A prototypical  model for  the study of  the influence of  disorder in
such an  ``athermal case'' is the  Gaussian RFIM. This  model has been
studied using two different approaches.   On the one hand, a number of
studies\cite{Koiller2000,Roters1999} have focused on the analysis of a
single magnetic interface.  The  numerical algorithms assume that only
spins close  to the  interface can flip.   On the other  hand, studies
including  both  nucleation  events  and interface  motion  have  been
performed     by     using     synchronous     relaxation     dynamics
\cite{Sethna1993,Sethna2001}.  In  both cases hysteresis  appears as a
consequence  of  the  local  fields  that keep  the  magnetic  moments
unreversed (pinned)  even at negative  values of the  applied external
field.   We will  focus on  the statistical  analysis of  such pinning
fields in the case of RFIM with synchronous relaxation dynamics.

Our  goal is  to point  out an  essential difference  between  the two
approaches.   The quenched  pinning fields  originating  from disorder
exhibit very  different statistical distributions in  both cases.  The
distribution  of  pinning fields  in  an  intermediate  state is  very
different from the quenched disorder distribution corresponding to the
initial  saturated state  (Gaussian)  in lattice  models that  include
nucleation and interface movement. This is due to the fact that in the
initial  stages of the  demagnetization process  the regions  with low
energy barriers have already  been reverted.  Close to coercivity, the
remaining  barriers are  much  higher due  to  the previous  selection
process.   Such  an effect  does  not occur  for  the  models with  an
advancing  single  interface, for  which  the  pinning  fields in  the
unreversed  regions always  exhibit a  statistical  distribution which
corresponds to the original quenched disorder distribution.

\section{Model}
\label{Model}
The 3d-Gaussian  RFIM at $T=0$ is  defined on a cubic  lattice of size
$N=L^3$.  On each lattice site a variable $S_i=\pm 1$ accounts for the
magnetic degrees of freedom. The Hamiltonian is:
\begin{equation}
H =- \sum_{ij} S_i S_j - \sum_{i} S_i h_i - B \sum_{i} S_i
\end{equation}
where  the first  term  stands for  the  exchange interaction  between
nearest-neighbour  spins $S_i$,  the second  term for  the interaction
with the quenched local random fields  $h_i$ and the last term for the
interaction  with  the driving  field  $B$.   The  $h_i$ are  Gaussian
distributed  with zero  mean and  variance $\sigma^2$.   Although this
model  enable  long-range  dipolar  forces  to  be  included,  from  a
computational  point  of  view  the numerical  solution  becomes  much
harder. Since we are interested  in the analysis of the pinning fields
generated by disorder, we have neglected long-range terms.

The  numerical  simulations   are  performed  using  local  relaxation
dynamics \cite{Sethna1993}.  The initial  saturated state with all the
spins $S_i=1$  corresponds to the equilibrium  state with $B=+\infty$.
The field  $B$ is  then decreased until  a spin $S_i$  becomes locally
unstable.  At this  point the external field is  kept constant and the
spin $S_i$ is flipped.  This may  cause an avalanche since some of its
neigbouring spins may become unstable.  All unstable spins are flipped
synchronously,  until the avalanche  ends. The  external field  $B$ is
then decreased again until a new avalanche starts.

\begin{figure}[p]
\begin{center}
\includegraphics[clip,width=12cm]{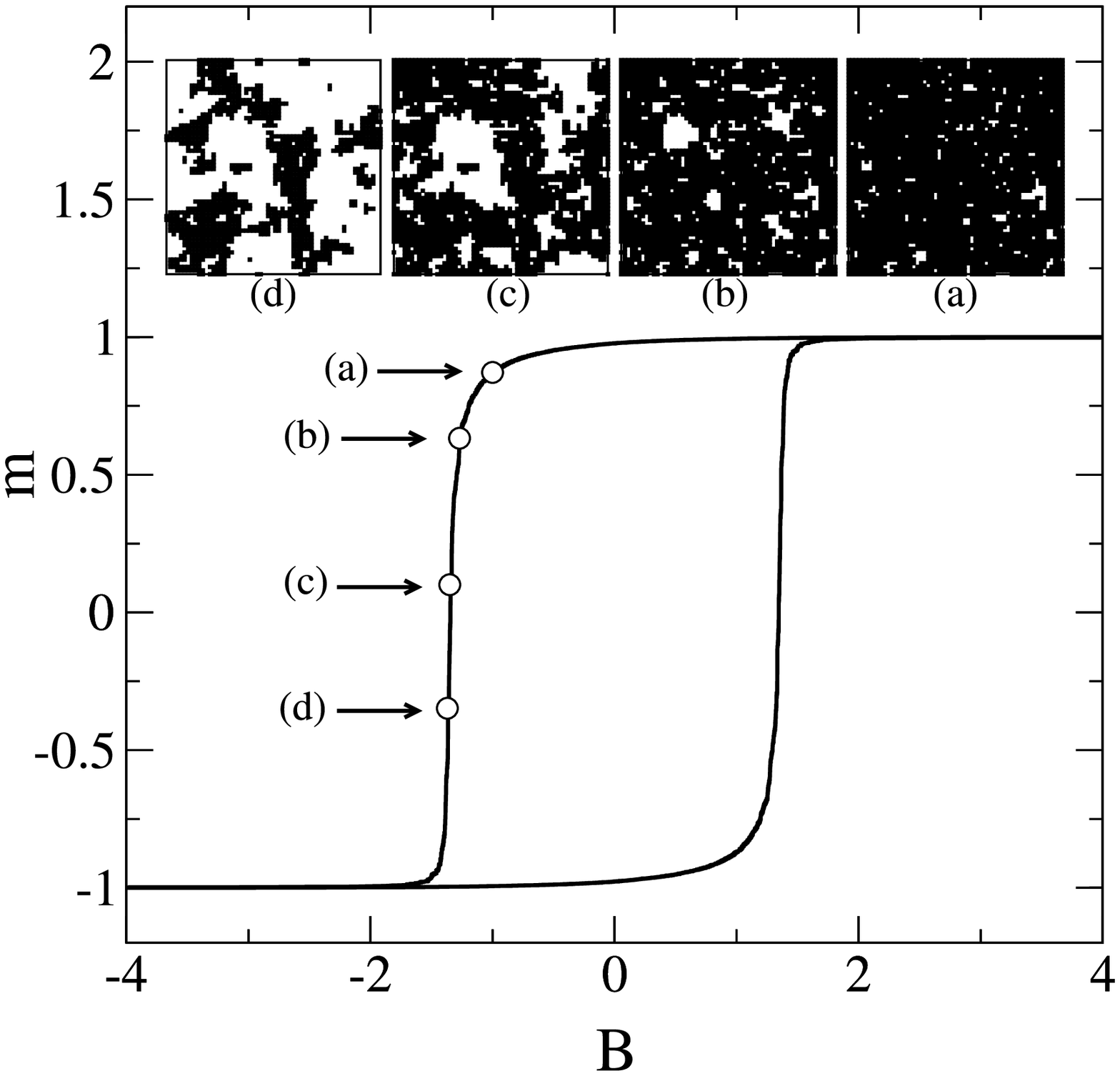}
\end{center}
\caption{Example  of a hysteresis  loop corresponding  to $\sigma=2.5$
and $L=60$.  The insets show examples of typical spin configurations.}
\label{fig1}
\end{figure}

The  hysteresis  loop  is  obtained  by  measuring  the  magnetization
$m=\sum_{i=1}^{N} S_i  / N$  as a function  of $B$.   Fig.  \ref{fig1}
shows an  example and the corresponding  configurations snapshots.  As
can  be  seen,  nucleation   and  interface  movement  coexist  during
evolution. Due to  the finite size of the  simulated system, the loops
consist of  a sequence of  discontinuous jumps or avalanches  for each
realization  of the  random  fields  $h_i$.  As  has  been studied  in
previous works \cite{PerezReche2003,Perkovic1999}, the characteristics
of  the  loops  depend on  the  amount  of  disorder $\sigma$  in  the
thermodynamic limit  ($L\rightarrow \infty$): they  are continuous for
$\sigma>\sigma_c$, whereas they display a discontinuity (corresponding
to a spanning avalanche)  at the coercive field for $\sigma<\sigma_c$.
This behaviour  is associated with  the existence of  a ``metastable''
critical  point   on  the   $(B,\sigma)$  phase  diagram   located  at
$\sigma_c=2.21$  and  $B_c=-1.425$.    The  behaviour  close  to  this
critical point can  be described by a set  of critical exponents.  For
instance, the correlation length diverges with an exponent $\nu \simeq
1.2$, the order parameter (the  magnetization jump $\Delta m$) goes to
zero with an  exponent $\beta \simeq 0.024 $  when $\sigma \rightarrow
\sigma_c$  from  below and,  at  $\sigma=\sigma_c$, the  magnetization
behaves as $m \sim |B-B_c|^{1/\delta}$ with $\delta \simeq 50 $.  Such
critical    exponents   have   been    obtained   by    detailed   FSS
analysis\cite{PerezReche2003}  which has also  revealed that  the most
convenient scaling  variable that measures the  distance to $\sigma_c$
is
\begin{displaymath}
u  =\frac{\sigma -  \sigma_c}{\sigma_c}  + A  \left  ( \frac{\sigma  -
\sigma_c}{\sigma_c} \right )^2
\end{displaymath}
with  $A=-0.2$.   Furthermore, since  we  will  be  interested in  the
measurement of properties as a  function of the external field $B$, we
will need a second scaling  variable to measure the distance to $B_c$.
The first simpler choice is:
\begin{equation}
v=\frac{B - B_{coe}}{B_{coe}}
\end{equation}
where $B_{coe}(\sigma,L)$ is the  coercive field that tends to $B_{c}$
when $\sigma\rightarrow \sigma_c$ and $L\rightarrow \infty$.

\section{Results}
\label{Results}

We define pinning fields $h_i^+$ as those quenched random fields $h_i$
for  which  $S_i=+1$  during   the  reversal  process  for  a  certain
intermediate  configuration. Although  the  set of  pinning fields  is
simply  a  subset  of  the  original  quenched  random  fields,  their
statistical  properties depend  on  the exact  path  followed until  a
certain configuration is reached.   We have computed the average value
of  the pinning  fields  $\langle h_i^+  \rangle$  and the  histograms
corresponding  to  their  statistical  distribution  $f(h_i^+)$  as  a
function  of  $B$ and  $\sigma$  by  simulating  many realizations  of
disorder.    Moreover   we   have   measured  pair   correlations   as
$C(h_i^+,h_j^+)=  L^3 \left  (  \langle h_i^+  h_j^+ \rangle-  \langle
h_i^+ \rangle^2 \right )$.

\begin{figure}[p]
\begin{center}
\includegraphics[clip,width=12cm]{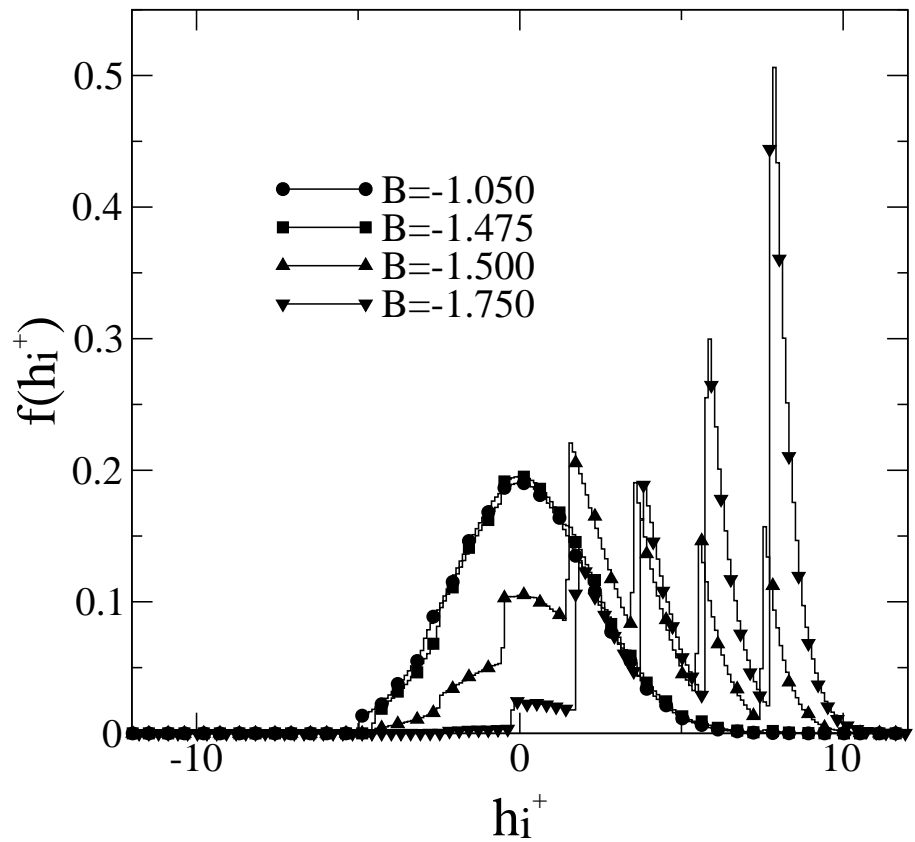}
\end{center}
\caption{Examples of distributions $f(h_i^+)$ corresponding to $L=60$,
$\sigma=2.14$  and   different  values   of  the  external   field  as
indicated. Data  corresponds to  averages over $1000$  realizations of
disorder.}
\label{fig2}
\end{figure}

Fig.  \ref{fig2}  shows the evolution  of the distribution  of pinning
fields along  the decreasing branch of a  hystereis loop corresponding
to $\sigma=2.14<\sigma_c$.  This distribution can be understood as the
distribution of  barriers created by the quenched  disorder that keeps
the   spins  in   the  metastable   state.   Although   initially  the
distribution  of pinning fields  is similar  to the  original Gaussian
distribution,  as the  magnetization decreases,  $f(h_i^+)$  starts to
develop a non-trivial structure.  In general the distribution tends to
shift to the right, towards the region of large pinning fields, but it
also develops  a number  of peaks associated  with the  seven possible
local magnetization environments.

\begin{figure}[p]
\begin{center}
\includegraphics[clip,width=12cm]{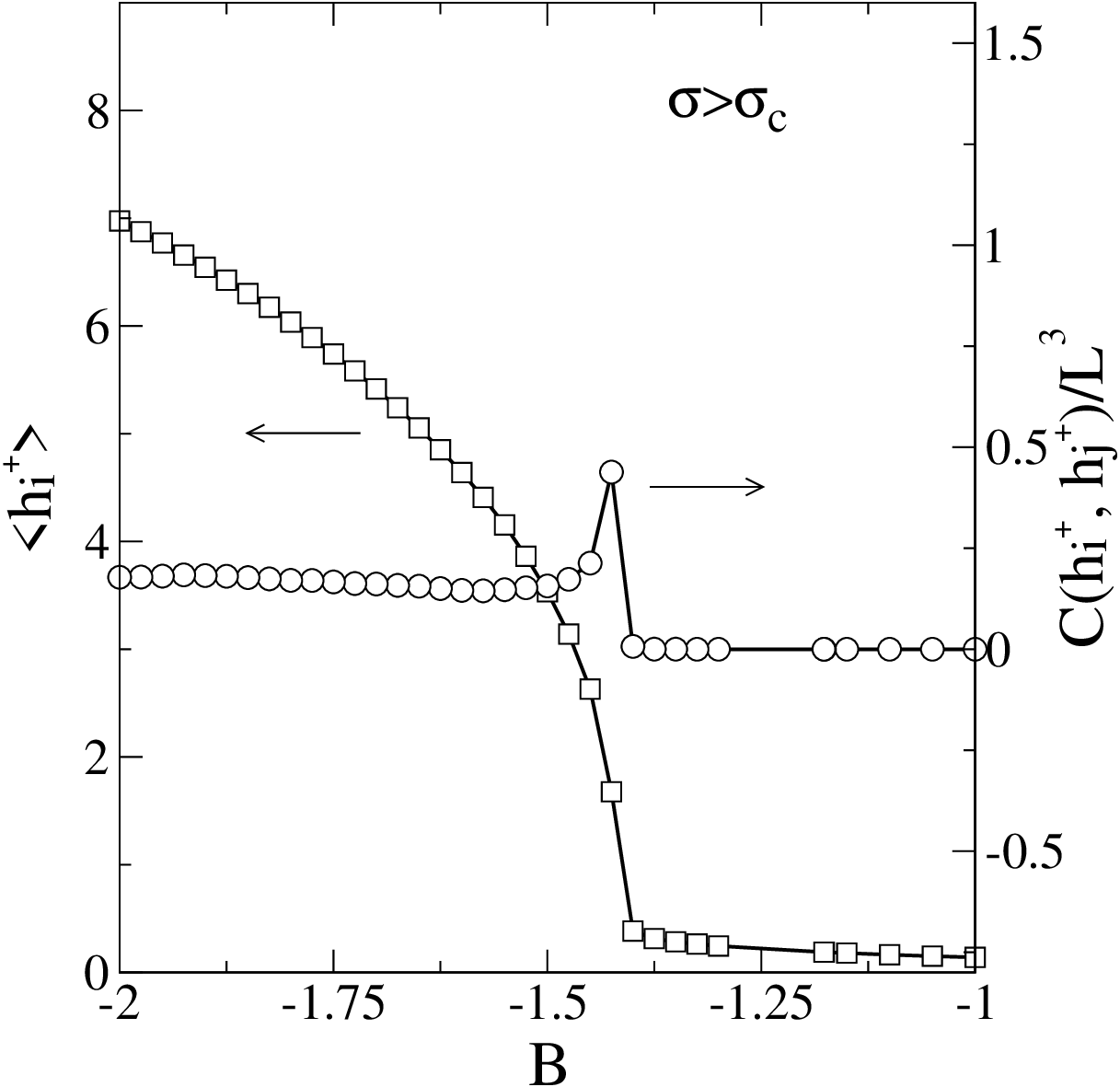}
\end{center}
\caption{Example of  the evolution of $\langle h_i^+  \rangle$ and $C(
h_i^+,h_j^+)$ as  a function of the  external field for  a system with
$L=60$  and $  \sigma=2.28$.  Data correspond  to  averages over  1000
realizations of disorder. }
\label{fig3}
\end{figure}

\begin{figure}[p]
\begin{center}
\includegraphics[clip,width=12cm]{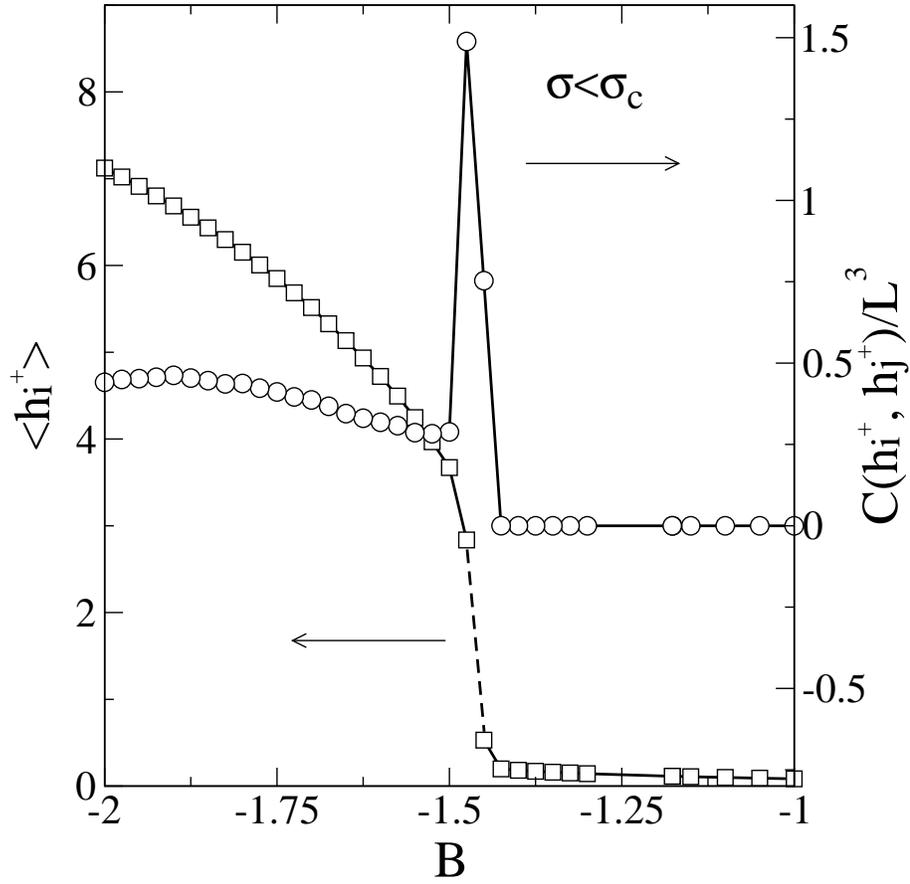}
\end{center}
\caption{Example of  the evolution of $\langle h_i^+  \rangle$ and $C(
h_i^+,h_j^+)$ as  a function of the  external field for  a system with
$L=60$  and  $\sigma=2.14$.  Data  correspond to  averages  over  1000
realizations of disorder.}
\label{fig4}
\end{figure}

Figs.   \ref{fig3} and  \ref{fig4} display  the evolution  of $\langle
h_i^+ \rangle$  and $C(  h_i^+,h_j^+)$ as a  function of  the external
field $B$  for two different  values of $\sigma$ corresponding  to two
cases: above and below $\sigma_c$. It is interesting to point out that
the correlation is not zero in the two cases and displays a peak close
to the  coercive field.  Note  that this means that  the distributions
$f(h_i^+)$ are only projections of complex multivariate distributions.
Additionally,  $\langle h_i^+  \rangle$ displays  a  discontinuity for
$\sigma<\sigma_c$.   The  behaviour  of  $\langle h_i^+  \rangle$  is,
therefore, similar  to the  behaviour of an  order parameter  around a
critical point.

\begin{figure}[p]
\begin{center}
\includegraphics[clip,width=13cm]{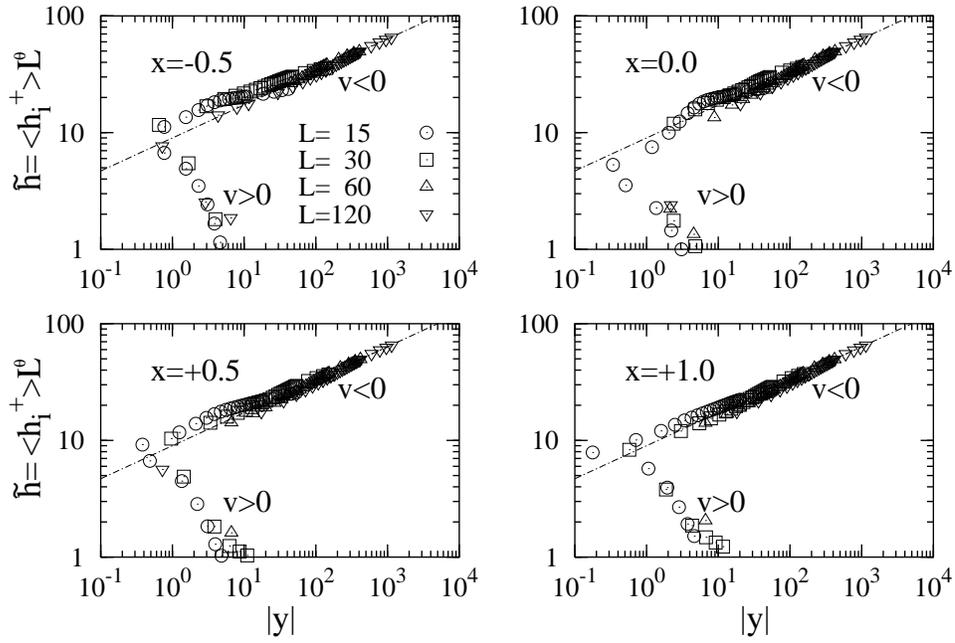}
\end{center}
\caption{FSS   analysis  of   the  average   pinning   field  $\langle
h_i^+\rangle$.   The  figures show,  on  log-log  scales, the  scaling
function  $\tilde h$ in  front of  $y=vL^{\beta \delta  / \nu}$  for 4
different values  of $x=uL^{1/\nu}$ as indicated on  each graph.  Data
correspond to  the overlap of sizes $L=15,30,60,120$  (as indicated by
the legend)  and averages over  many realizations of  disorder ranging
from $10^6$ to $10^3$.}
\label{fig5}
\end{figure}

\begin{figure}[p]
\begin{center}
\includegraphics[clip,width=12cm]{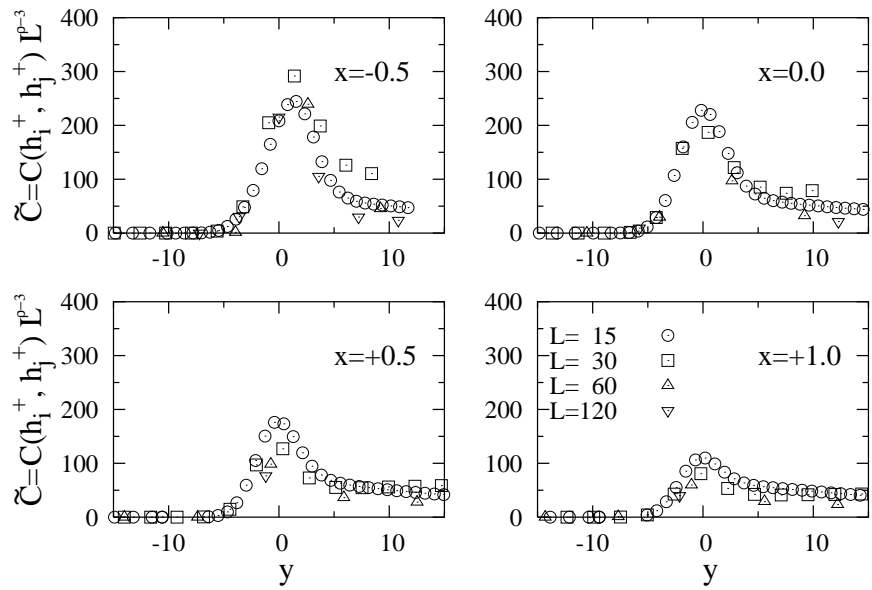}
\end{center}
\caption{FSS  analysis  of  the  correlation  between  pinning  fields
$C(h_i^+,h_j^+)$. Details are the same as in Fig.~\ref{fig5}.}
\label{fig6}
\end{figure}

Quantitative analysis of such  data requires a convenient FSS analysis
to be carried out. The two analyzed properties $\langle h_i^+ \rangle$
and $C(  h_i^+,h_j^+)$ are  functions of the  external field  $B$, the
amount of disorder  $\sigma$ and the system size  $L$.  The FSS ansatz
allows  to   the  singular  (critical)  contributions   of  these  two
properties  to   be  expressed  as   a  function  of   the  invariants
$x=uL^{1/\nu}$ and $y=vL^{\beta \delta / \nu} $ :
\begin{equation}
\langle    h_i^+   \rangle=   L^{-\theta}    {\tilde   h}    \left   (
uL^{1/\nu},vL^{\beta \delta / \nu} \right )
\end{equation}
\begin{equation}
C(h_i^+,h_j^+)=  L^{3-\rho}  {\tilde  C}  \left  (uL^{1/\nu},vL^{\beta
\delta / \nu} \right )
\end{equation}
The exponents  $\theta$ and  $3-\rho$ characterize how  $\langle h_i^+
\rangle$  decreases and  how  the correlation  $C$ diverges.   Figures
\ref{fig5} and \ref{fig6} show  the scaling functions ${\tilde h}$ and
${\tilde C}$. The good quality  of the collapses of data corresponding
to  different system  sizes,  confirms the  scaling assumptions.   The
exponents that allow  the best collapses are $\theta  \simeq 0.41$ and
$3-\rho  \simeq  1.80$.   Moreover,   the  behaviour  of  the  scaling
functions allows prediction  of how the average pinning  field and the
correlations  behave in  the  thermodynamic limit.   Since $\tilde  h$
behaves as $y^{\theta \nu /  \beta \delta}$ for $v<0$ (as indicated by
the  discontinuous  line)  ,  $\langle  h_i^+\rangle$  is  finite  for
$B<B_c$.  The scaling  behaviour for $v>0$ is not  so good since there
are  non-scaling contributions  due to  the existence  of non-spanning
avalanches  for  all  values  of $\sigma$  \cite{PerezReche2003}.   As
regards  correlations, the  scaling functions  display a  peak profile
which   indicates  that,   besides   non-scaling  contributions,   the
correlation   $C(h_i^+,h_j^+)$    will   diverge   at    $B=B_c$   and
$\sigma=\sigma_c$ in the thermodynamic limit.

\section{Conclusions}
\label{Conclusions}

Pinning fields are  responsible for the energy barriers  that keep the
spins in  the metastable state  within the context  of the 3d  RFIM at
$T=0$.   The statistical  distribution  of pinning  fields during  the
reversal magnetization process has  been studied.  Initially, when the
system  is saturated,  the  pinning fields  are trivially  distributed
according to  the nominal Gaussian distribution of  random fields.  As
the demagnetizing  process advances and domains of  negative spins are
created,  the pinning  fields display  a complex  distribution.  Their
mean value  increases monotonously for decreasing $B$.   For $\sigma >
\sigma_c$ this increase is continuous  but for $\sigma < \sigma_c$ the
average  pinning   field  displays  a   discontinuity  at  coercivity.
Moreover we  have shown  that close to  coercivity the  pinning fields
exhibit strong statistical correlations. We finally remark that such a
complex behaviour of  the distribution of pinning fields  is not taken
into account in the studies that focus on the analysis of an advancing
single magnetic interface.

We acknowledge fruitful  discussions with F.J.P\'erez-Reche. This work
has received  financial support  from CICyT (Project  No MAT2001-3251)
and CIRIT (Project  2001SGR00066).  X.I. aknowledges financial support
from DGI-MCyT (Spain).


\end{document}